\documentstyle[preprint,aps]{revtex}
\begin{document}
%\setlength{\baselineskip}{12pt}
%\begin{flushright}
%\end{flushright}
%\vspace{2cm}
\draft
\newfont{\form}{cmss10}
\newcommand{\unity}{1\kern-.65mm \mbox{\form l}}
\newcommand{\k}{\mbox{\form l}\kern-.6mm \mbox{\form K}}
\newcommand{\D}{D \raise0.5mm\hbox{\kern-2.0mm /}}
\newcommand{\A}{A \raise0.5mm\hbox{\kern-1.8mm /}}
\def\pmb#1{\leavevmode\setbox0=\hbox{$#1$}\kern-.025em\copy0\kern-\wd0
\kern-.05em\copy0\kern-\wd0\kern-.025em\raise.0433em\box0}
\title{Ghost decoupling in 't Hooft spectrum for mesons}
\author{A. Bassetto}
\address{Dipartimento di Fisica ``G.Galilei", Via Marzolo 8 -- 35131
Padova, Italy \\ INFN, Sezione di Padova, Italy}
\author{L. Griguolo}
\address{Center for Theoretical Physics, Laboratory of
Nuclear Science, MIT, Cambridge, MA 02139, USA}

\date{\today}
\maketitle
\begin{abstract}
We show that the
replacement of the ``instantaneous'' 't Hooft's potential with the
causal form suggested 
by equal time canonical
quantization in light-cone gauge, which entails the occurrence of
negative probability states, does not change the bound state 
spectrum when the difference is treated as a single insertion in the
kernel.
\end{abstract}
\vfill\eject

\narrowtext
In 1974 G. 't Hooft \cite{Ho74} proposed a very interesting model to describe
the mesons, starting from a SU(N) Yang-Mills theory in 1+1 dimensions
in the large N limit.

Quite remarkably in this model quarks look confined, while a discrete
set of quark-antiquark bound states emerges, with squared masses lying
on rising Regge trajectories.

The model is solvable thanks to the ``instantaneous'' character of
the potential acting between quark and antiquark.

Three years later such an approach was criticized by T.T. Wu \cite{Wu77},
who replaced the instantaneous 't Hooft's potential by an expression
with milder analytical properties, allowing for a Wick's rotation
without extra terms.

Unfortunately this modified formulation led to a quite involved bound
state equation, which may not be solved. An attempt to treat it
numerically in the zero bare mass case for quarks \cite{Be78} led only to
partial answers in the form of a completely different physical
scenario. In particular no rising Regge trajectories were found.

After those pioneeristic investigations, many interesting papers 
followed 't Hooft's approach, pointing out further remarkable
properties of his theory and blooming into the recent achievements
of two dimensional QCD , whereas Wu's approach sank into oblivion,
if not disrepute.

Still, equal time canonical quantization of Yang-Mills theories
in light-cone gauge \cite{Da85} leads precisely in 1+1 dimensions
to the Wu's expression for the 
vector exchange between quarks \cite{Ba94}, which is nothing but the 1+1
dimensional version of the Mandelstam-Leibbrandt (ML) \cite{Da85,Ml84} 
propagator.
Its causal nature, which entails the occurrence of
negative probability states, makes it compulsory in order to achieve gauge 
invariance and renormalization in 1+(D-1) dimensions \cite{Ko93,Ba91}.

Purpose of this note is to show that indeed the difference between
't Hooft's ${{1}\over {k_{-}^2}}$ and Wu's ${{1}\over {(k_{-}-i\epsilon
sign (k_{+}))^2}}$ potentials might in some sense be treated
as a perturbation. Our main result will be that no correction
due to this difference will
affect the 't Hooft's bound state spectrum in the small 
coupling region, owing to a precise cancellation between ``real'' and 
``virtual'' insertions. This phenomenon is analogous to the one occuring,
with respect to the same extra term, in perturbative four-dimensional 
calculations concerning
Altarelli-Parisi  \cite{Ba93} and Balitsky-Fadin-Kuraev-Lipatov 
\cite{Ry93} kernels. This analogy may have far-reaching consequences.

We follow here definitions and notations of 
refs.\cite{Ho74} and \cite{Wu77} the reader is
invited to consult.

The 't Hooft potential exhibits an infrared singularity
which, in the original formulation, was handled by introducing
an infrared cutoff; a quite remarkable feature of this theory
is that bound state wave functions and related eigenvalues
turn out to be cutoff independent. As a matter of fact in
ref. \cite{Ca76}, it has been pointed out that the singularity
at $k_{-}=0$ can also be regularized by a Cauchy principal
value ($P$) prescription without finding differences in gauge
invariant quantities. Then, the difference between the two
potentials is represented by the following distribution

\begin{equation}
\label{unoa}
\Delta (k)\equiv {{1}\over {(k_{-}-i\epsilon
sign (k_{+}))^2}} - P\Big({{1}\over {k_{-}^2}}\Big)= - i \pi 
sign (k_{+}) \delta^{\prime}(k_{-}).
\end{equation}

This is the quantity we are going to treat as an insertion
in the Wu's 
integral
equations for the quark propagator and for the bound state wave
function, starting from 't Hooft's solutions. We stress that we
shall sum exactly the same planar diagrams of 
refs.\cite{Ho74} and \cite{Wu77}, which are the relevant ones
in the large $N$ limit.

The Wu's integral equation for the quark self-energy in the Minkowski
momentum space is

\begin{eqnarray}
\label{unob}
\Sigma(p;\eta)&=& i {{g^2}\over {\pi^2}} {{\partial}\over {\partial p_{-}}}
\int dk_{+}dk_{-} \Big[P\Big({{1}\over {k_{-}-p_{-}}}\Big)+
i \eta \pi sign (k_{+}-p_{+}) \delta (k_{-}-p_{-})\Big]\nonumber\\
&\cdot&{{k_{-}}\over {k^2+m^2-k_{-}\Sigma (k;\eta)-i\epsilon}},
\end{eqnarray}

where $\eta$ is a real parameter which will be used in the sequel
as a counter of insertions and here should be set equal to 1.

Its exact solution with appropriate boundary conditions reads

\begin{eqnarray}
\label{uno}
\Sigma(p;\eta)&=& {{1}\over {2p_{-}}}\Big(\Big[p^2+m^2+(1-\eta){{g^2}\over 
{\pi}}\Big]-\Big[p^2+m^2-(1-\eta){{g^2}\over 
{\pi}}\Big]\nonumber\\
&\cdot&\sqrt {1- {{4\eta g^2 p^2}\over {\pi(p^2+m^2-(1-\eta){{g^2}\over 
{\pi}}}-i\epsilon)^2}}\Big).
\end{eqnarray}

One can immediately realize that 't Hooft's and Wu's solutions
are recovered for $\eta =0$ and $\eta =1$ respectively.

The dressed quark propagator turns out to be

\begin{equation}
\label{due}
S(p;\eta) = - {{i p_{-}}\over {m^2+2 p_{+}p_{-}- p_{-}\Sigma(p;\eta)}}. 
\end{equation}

Wu's bound state equation in
Minkowski space, using light-cone coordinates, is

\begin{eqnarray}
\label{tre}
\psi(p,r)&=& {{-ig^2}\over {\pi ^2}} S(p;\eta) S(p-r;\eta)
\int dk_{+}dk_{-} \Big[P\Big({{1}\over 
{(k_{-}-p_{-})^2}}\Big)+\nonumber\\
&+&
i \eta \pi sign (k_{+}-p_{+}) \delta^{\prime} (k_{-}-p_{-})\Big] 
\psi(k,r). 
\end{eqnarray}                                                            

We are here considering for simplicity the equal mass case and $\eta$
should be set equal to 1.

Let us denote by $\phi_{k}(x),\,\, 0\le \! x= {{p_{-}}\over {r_{-}}}\le 
\!1,\,\, r_{-}>0$, 
the 't Hooft's eigenfunction corresponding
to the eigenvalue $\alpha_{k}$ for the quantity ${{-2 r_{+}r_{-}}\over 
{M^2}}$, where $M^2= m^2 - {{g^2}\over {\pi}}$. 
Those eigenfunctions are
real, of definite parity under the exchange $x \to 1-x$ and vanishing 
outside the interval $0<x<1$:

\begin{eqnarray}
\label{quattro}
\phi_{k}(x)&=& \int dp_{+} {{r_{-}}\over {M^2}} 
\psi_{k}(p_{+},p_{-},r),\nonumber\\
\psi_{k}&=& {{1}\over {i \pi}}\phi_{k}(x) {{M^4}\over {M^2+2r_{-}p_{+}x 
-i\epsilon}}\,\,{{1-\alpha_{k}x(1-x)}\over {M^2-\alpha_{k}M^2(1-x)-
2r_{-}p_{+}(1-x)- i \epsilon}},
\end{eqnarray}                                                            

They  are solutions of eq.(\ref{tre}) for $\eta=0$ and
form a complete set.

We are interested in a first order calculation in $\eta$.
Of course this procedure
is to be considered in a heuristic way;
moreover it is likely to be sensible only in the weak coupling region
${{g^2}\over {\pi}}< M^2$.
The integral equation (\ref{tre}), after first order 
expansion in $\eta$ of its kernel, becomes

\begin{eqnarray}
\label{cinque}
\psi(p_{+},p_{-},r)&=& {{ig^2}\over {\pi^2}}{{p_{-}}\over 
{M^2+2p_{+}p_{-}-i\epsilon}}{{p_{-}-r_{-}}\over 
{M^2+2(p_{+}-r_{+})(p_{-}-r_{-})-i\epsilon}}\nonumber\\
&\cdot&\Big[\Big(1-{{\eta g^2 M^2}\over 
{\pi}}[(M^2+2p_{+}p_{-}-i\epsilon)^{-2}+(M^2+2(p_{+}-r_{+})(p_{-}-r_{-})
-i\epsilon)^{-2}]\Big)\nonumber\\
&\cdot&\int dk_{+}dk_{-} 
P{{1}\over{(k_{-}-p_{-})^2}}\psi(k_{+},k_{-},r)-\nonumber\\
&-&i \pi \eta \int dk_{+}dk_{-} sign(k_{+}-p_{+})
\delta^{\prime}(k_{-}-p_{-})\psi(k_{+},k_{-},r)\Big].
\end{eqnarray}                                                            

We integrate this equation over $p_{+}$ with $r_{-}>0$ and search
for solutions with the same support properties of 't Hooft's ones.
We get

\begin{eqnarray}
\label{cinquea}
\phi(x,r)&=& {{g^2}\over{\pi M^2}}{{x(1-x)}\over{1-\alpha 
x(1-x)-i\epsilon}}\Big[\Big(1-\eta{{g^2}\over{\pi M^2}}
{{x^2+(1-x)^2}\over{(1-\alpha 
x(1-x)-i\epsilon)^2}}\Big)\nonumber\\
&\cdot& P\int_0^1 {{dy}\over {(y-x)^2}}\phi(y,r)
-{{\alpha \eta}\over{2}}\int d\xi \log {{{{1}\over{1-x}}-\alpha (1-\xi)-i
\epsilon}\over {{{1}\over{x}}-\alpha 
\xi-i\epsilon}}\psi^{\prime}(\xi,x,r)\Big],
\end{eqnarray}

where $\prime$ means derivative with respect to $x$.

It is now straightforward to check that 't Hooft's solution 
$\psi_{k}(p_{+},p_{-},r)$ is indeed a solution also of this 
equation when $\alpha$
is set equal to $\alpha_{k}$, for any value of $\eta$, in particular
for $\eta=1$, thanks to a precise cancellation of the contributions
coming from the propagators (``virtual'' insertions) against the
extra term due to the modified form of the ``potential'' (``real''
insertion). In other words the extra piece of the kernel at
$\alpha=\alpha_{k}$ vanishes when acting on $\psi_{k}$.

As a matter of fact, taking 't Hooft's equation
into account, we get

\begin{eqnarray}
\label{sei}
&&(\alpha_{k}-\alpha)\phi_{k}(x)\Big[1-{{\eta g^2}\over{\pi M^2
[1-\alpha x(1-x)-i\epsilon]^2}}\Big((1-x)^2
+x^2[1+{{1-\alpha x(1-x)}\over{1-\alpha_{k} 
x(1-x)-i\epsilon}}]\Big)\Big]=\nonumber\\
&=& {{\eta g^2}\over {\pi 
M^2}}\,\,\phi_{k}^{\prime}(x)\,\, log {{1-\alpha_{k}x(1-x)-i\epsilon}
\over{1-\alpha x(1-x)-i\epsilon}}.
\end{eqnarray}                                                            

There are no corrections from a single insertion in the kernel to
't Hooft eigenvalues and eigenfunctions.
We stress that this result does
not depend on their detailed form, 
but only on their general properties.

\vskip .5truecm

In conclusion the ghosts which are responsible of the causal
behaviour of the ML propagator \cite{Ba85}, do not modify
the bound state spectrum, as their ``real'' contribution 
cancels against the ``virtual'' one in propagators.
Wu's equation for colorless bound states, 
although much more involved than the
corresponding 't Hooft's one, might still apply.
This is at least
the heuristic lesson one learns from a single insertion in the kernel
and is in agreement with a similar mechanism occuring
in four-dimensional perturbative QCD \cite{Ba93,Ry93}. 

Higher order insertions, although worth to be
explored, do not lead to a trivial problem owing to the non linear
character of the integral equation for the propagator.
Moreover the validity of our result is limited to the small coupling
region; for ${{g^2}\over {\pi M^2}}>1$
the theory is likely to be in a different phase (see for instance
\cite{Zy95})
and Wu's equation may lead to a quite different physical scenario.

In QCD$_{4}$ consistency with renormalization procedure for general
Green's functions strongly suggests the ML prescription to regularize
infrared singularities; in QCD$_{2}$ the ML option may not be strictly
compelling as the theory becomes super-renormalizable. 

Planarity plays a crucial role in our considerations; as a matter 
of fact the same ghost contribution cancels also in Wilson loops
provided the large $N$ limit is considered \cite{Cy95}. Indeed
cross diagrams are order ${{1}\over{N^2}}$ with respect to planar ones,
which in turn are unaffected by the ML prescription.
Planarity and ghost cancellation 
might be deeply related, an argument which is worth of further 
investigation.
\vskip .5truecm

ACKNOWLEDGEMENTS.
We thank G. Nardelli for many useful discussions. L.G. acknowledges
a INFN post-doctoral fellowship.


\begin{thebibliography}{100}              

\bibitem{Ho74}{ G. 't Hooft, Nucl. Phys. {\bf B75}, 461 (1974).}
\bibitem{Wu77}{ T.T. Wu, Phys.Lett. {\bf 71B}, 142 (1977).}
\bibitem{Be78}{ N.J. Bee, P.J. Stopford, B.R. Webber, Phys.Lett. 
{\bf 76B}, 315 (1978).}
\bibitem{Ba93}{A. Bassetto in ``QCD and High Energy Hadronic 
Interaction'',
ed. J. Tran Thanh Van, pg. 103 , Editions Frontieres 1993.}
\bibitem{Ry93}{A. Bassetto, M. Ryskin, Phys. Lett. {\bf B316}, 542 
(1993).}
\bibitem{Ca76}{C.G. Callan, N. Coote, D.J. Gross, Phys. Rev. {\bf D13},
1649 (1976).}
\bibitem{Da85}{A. Bassetto, M. Dalbosco, I. Lazzizzera, R. Soldati,
Phys. Rev. {\bf D31}, 2012 (1985).}
\bibitem{Ba94}{ A. Bassetto, F. De Biasio and L.Griguolo, 
Phys.Rev.Lett. {\bf 72}, 3141 (1994).}
\bibitem{Ml84}{ S. Mandelstam, Nucl. Phys. {\bf B213}, 149 (1983); G. 
Leibbrandt, Phys. Rev. {\bf D29}, 1699 (1984).}
\bibitem{Ko93}{ A. Bassetto, I.A. Korchemskaya, G.P. Korchemsky and 
G. Nardelli, Nucl.Phys. {\bf B408}, 52 (1993).}
\bibitem{Ba91}{A. Bassetto, G. Nardelli 
and R. Soldati,
Yang-Mills theories in algebraic non-covariant gauges, (World Scientific,
Singapore, 1991) and references therein.}
\bibitem{Ba85}{ A. Bassetto, Proceedings of the VIII Warsaw Symposium
on Elementary Particle Physics - Kazimierz (Poland)1985.}
\bibitem{Zy95}{B. Chibisov, A.R. Zhitnitsky, Phys. Lett. {\bf B362},
105 (1995).}
\bibitem{Cy95}{A. Bassetto, L. Griguolo, G. Nardelli, preprint hep-th
9506095.}
\end{thebibliography}
\end{document}